\documentclass[preprint2]{aastex}

\shorttitle{37 GHz observations of BL Lacs}
\shortauthors{Nieppola et al.}

\begin{document}

\title{37 GHz observations of a large sample of BL Lacertae objects}

\author{Elina Nieppola\altaffilmark{1,2}, Merja Tornikoski\altaffilmark{1}, Anne L\"ahteenm\"aki\altaffilmark{1}, Esko Valtaoja\altaffilmark{2,3}, Tero Hakala\altaffilmark{1}, Talvikki Hovatta\altaffilmark{1}, Mikko Kotiranta\altaffilmark{1}, Sakari Nummila\altaffilmark{2}, Tapio Ojala\altaffilmark{2}, Mikko Parviainen\altaffilmark{1}, Mikko Ranta\altaffilmark{2}, Pia-Maria Saloranta\altaffilmark{1,2}, Ilona Torniainen\altaffilmark{1}, Mirko Tr\"oller\altaffilmark{1}}

\altaffiltext{1}{Mets\"ahovi Radio Observatory, Mets\"ahovintie 114, 02540 Kylm\"al\"a, Finland; elina.nieppola@tkk.fi}
\altaffiltext{2}{Tuorla Observatory, V\"ais\"al\"antie 20, 21500 Piikki\"o, Finland}
\altaffiltext{3}{Dept. of Physical Sciences, University of Turku, 20100 Turku, Finland}

\begin{abstract}
We present 37 GHz data obtained at Mets\"ahovi Radio Observatory in 2001 December - 2005 April for a 
large sample of BL Lacertae objects. We also report the mean variability indices and radio spectral indices in frequency intervals 5 - 37 GHz and 37 - 90 GHz.
Approximately 34 \% of the sample was detected at 37 GHz, 136 BL Lacertae objects in all. A large majority of 
the detected sources were low-energy BL Lacs (LBLs). The variability 
index values of the sample were diverse, the mean fractional variability of the sample being $\Delta S_2$=0.31. 
The spectral indices also varied widely, but the average radio spectrum of the sample sources is flat. 
Our observations show that many of the high-energy BL Lacs (HBL), which are usually considered radio-quiet, can at times be detected at 37 GHz. 
\end{abstract}


\defcitealias{nieppola06}{Paper I}
\defcitealias{veron00}{VCV2000}

\section{INTRODUCTION}

BL Lacertae objects (BL Lacs, BLOs) demonstrate the most violent
behaviour known among active galactic nuclei (AGN). Their basic properties include high optical
polarisation, rapid and strong variability, and a near-featureless
spectrum \citep[][ for an extensive review on AGN, see \citet{urry95}]{stein76, kollgaard94, jannuzi94}. The selection criteria in many surveys also include a flat
radio spectrum ($\alpha\leq0.5$, $S\propto\nu^{-\alpha}$) \citep[][ and the references therein]{padovani95_II}.

The spectral energy distributions (SEDs) of BL Lacertae objects are thought to consist of two components; 
a synchrotron component at lower frequencies and an inverse Compton (IC) component at higher frequencies. The locations 
of the components in the log\,$\nu$ -- log\,$\nu F_{\nu}$ plane varies much, and this has been used as a classification basis for BL Lacs 
\citep[][ hereafter Paper I]{padovani95_I, giommi95, sambruna96, nieppola06}. For low-energy peaked BL Lacs (LBLs) the peak frequency of the synchrotron component is at radio and IR-frequencies, for intermediate BL Lacs (IBLs) in the optical and UV-frequencies, and for high-energy peaked BL Lacs (HBLs) in the X-ray band. To further explore 
the physics behind this spectral 
sequence, it is important to know which are the properties that set LBLs and HBLs apart. 

Radio monitoring provides insight into the variability characteristics
and long-term behaviour of AGN. Mets\"ahovi is one of the few
observatories which target the high radio frequencies, and the only one observing at 37 GHz. BL Lacs have been
observed in Mets\"ahovi as part of the ongoing
Mets\"ahovi - Tuorla Observatory collaboration project. There are currently 398 BL Lacs in the Mets\"ahovi AGN sample.
We wanted to study this large sample of BL Lacs
at 37 GHz because many of them had not previously been studied
at high radio frequencies.
Our aim was to measure the flux densities of the various subsamples of BLOs 
in order to get a full understanding of the high radio frequency 
behaviour of these objects, all the way from the radio-selected BL 
Lacs (RBLs) to the X-ray selected BL Lacs (XBLs).
HBLs and XBLs have not been actively studied in the
high radio frequency domain.  Even though XBLs seem to be
significantly weaker at radio wavelengths than RBLs, there is
little evidence of radio-silent BLOs. \citet{stocke90} suggest that they do not exist, and even though there
have been some recent studies on possibly completely radio-silent BL
Lac candidates \citep{londish02, londish04}, the number
of such BLOs is probably low.  Because the only high
radio frequency study on XBLs is a very small sample studied by \citet{gear93}, and only
the low frequency radio flux densities of these sources are known in general,
it is possible that some of the rarely studied BLOs exhibit
``surprising'' spectral shapes in the high radio frequency domain,
such as the extreme inverted-spectrum sources \citep{tornikoski00_southern, tornikoski01_gps, torniainen05}.

Our studies also support the work of Extragalactic foreground sources 
Working Group of the Planck satellite\footnote{http://www.rssd.esa.int/Planck}, 
particularly that of the Low Frequency Instrument (LFI) Consortium. In preparation for the 
Planck mission, one of our tasks is to estimate the number of extragalactic sources 
detectable at the Planck frequencies. The existence
of unforeseen bright sources could seriously affect the primary task
of Planck, i.e., the mapping of the cosmic microwave background (CMB). The theoretical aspects of blazar contamination in CMB maps have been thoroughly studied by \citet{giommi04,giommi06,toffolatti98}. Because of the lack
of actual measurements at high radio frequencies, many of the
source statistics have been based on extrapolated low-frequency data.
Therefore we find it important to investigate, also in practice, if there are source populations or subpopulations that are brighter in
the high radio frequencies than earlier assumed, especially such that exhibit
significant variability and can at times make a significant
contribution to the contamination of the CMB maps. It is vital to
understand the high radio frequency behaviour of the various BLOs to
see how many of them, at least some of the time, can also be detected
by Planck.

The main objective of this paper is to publish the data we have
collected for almost three and a half years. We have also calculated some descriptive parameters for the sample. In \S2 we introduce the sample and describe the observations, in \S3 we state the detection rates by class. In \S4 and \S5 we report the
values of the variability indices and spectral indices we found for
the sample, and in \S6 say a few words about the variability
timescales. We finish with a discussion and conclusions in \S7 and \S8. All
statistical tests have been performed with the Unistat
5.0 software. 

\section{THE SAMPLE AND OBSERVATIONS}
\label{obs}

The Mets\"ahovi BL Lac sample consists of 398 objects, 382 of which are from the Veron-Cetty \& Veron BL Lac catalogue (Veron-Cetty \& Veron 2000, hereafter VCV2000), and 17 objects from the
literature. \citetalias{veron00} consists of sources from several different surveys with a
wide range of selection criteria. The Mets\"ahovi sample is one of the largest
BL Lac samples ever studied and offers a relatively unrestricted view of this
group of AGNs. The declination range covered by the Mets\"ahovi BL Lacs
is from $\mathrm{-11^o}$ to $\mathrm{86\,^o}$.

Most of the sample sources were classified as LBL/IBL/HBL in \citetalias{nieppola06}
based on their synchrotron peak frequencies,
$\nu_{peak}$, calculated from a parabolic fit to their SEDs. The class boundaries were log\,$\nu_{peak}\le$\,14.5 for LBLs,
14.5\,$<\,$log\,$\nu_{peak}\le$\,16.5 for IBLs and log\,$\nu_{peak}>$\,16.5 for HBLs. There are 98 LBLs, 96 IBLs and
110 HBLs in the sample. 94 sources remained unclassified on account of poor spectral fits. 
The full source sample is listed in Table~\ref{sample}. For most sources, we use the object naming convention adopted from \citetalias{veron00}. The subclass of each source is also listed. 

\begin{deluxetable}{lcccccccc}

\tablecaption{The Mets\"ahovi BL Lac sample.\label{sample}}
\tabletypesize{\small}
\tablewidth{0pt}
\tablehead{\colhead{Source name} & \multicolumn{3}{c}{Right Ascension} & \colhead{} & \multicolumn{3}{c}{Declination} & \colhead{class\tablenotemark{a}} \\ 
\cline{2-4} \cline{6-8}\\
\colhead{} & \colhead{(h)} & \colhead{(min)} & \colhead{(s)} & \colhead{} & \colhead{(deg)} & \colhead{(arcmin)} & \colhead{(arcsec)} & \colhead{} } 

\startdata
NRAO 5 & 0 & 6 & 13.9 & & -6 & 23 & 36 & LBL \\
RX J0007.9+4711 & 0 & 7 & 59.9 & & +47 & 12 & 7 & IBL \\
MS 0011.7+0837 & 0 & 14 & 19.7 & & +8 & 54 & 4 & HBL \\
RXS J0018.4+2947 & 0 & 18 & 27.8 & & +29 & 47 & 32 & \nodata\\
PKS 0017+200 & 0 & 19 & 37.9 & & +20 & 21 & 46 & LBL \\
\enddata

\tablecomments{Table~\ref{sample} is published in its entirety in the electronic edition of the Astronomical Journal. A portion is shown here for guidance regarding its form and content.}
\tablenotetext{a}{as determined in \citet{nieppola06}.}
\end{deluxetable}

\begin{deluxetable}{lcccccccc}

\tablewidth{0pt}
\tabletypesize{\footnotesize}
\tablecaption{37 GHz flux densities for the Mets\"ahovi BL Lac sample.\label{fluxes}}

\tablehead{\colhead{Source name} & \multicolumn{6}{c}{Date of the observation (J2000)} & \colhead{Flux density} & \colhead{Error} \\ 
\cline{2-7}\\
\colhead{} & \colhead{(year)} & \colhead{(month)} & \colhead{(day)} & \colhead{(h)} & \colhead{(min)} & \colhead{(Julian date)} & \colhead{(Jy)} & \colhead{(Jy)} } 

\startdata
NRAO 5 & 2002 & 5 & 24 & 6 & 40 & 2452418.778 & 1.71 & 0.17 \\
NRAO 5 & 2003 & 3 & 30 & 9 & 20 & 2452728.889 & 2.94 & 0.25 \\
NRAO 5 & 2005 & 2 & 9 & 13 & 10 & 2453411.049 & 2.69 & 0.16 \\
RX J0007.9+4711 & 2002 & 3 & 17 & 6 & 28 & 2452350.769 & $<$\,0.36 & \nodata \\
RX J0007.9+4711 & 2003 & 3 & 16 & 5 & 56 & 2452714.747 & $<$\,0.2 & \nodata \\
\enddata

\tablecomments{Table~\ref{fluxes} is published in its entirety in the electronic edition of the Astronomical Journal. A portion is shown here for guidance regarding its form and content. The error is listed for $S/N>$\,4 detections only; for non-detections we have calculated $S/N>$\,4 upper limits.}

\end{deluxetable}

Mets\"ahovi radio telescope is a radome enclosed
antenna with a diameter of 13.7 metres. It is situated in Kirkkonummi,
Finland, at 60 m above the sea level. It has been used for AGN variability studies at 22, 37 and 87 GHz.
The observations for this project were carried out using
the 37 GHz receiver, which is a dual horn, Dicke-switched receiver 
with a HEMT preamplifier operated at room temperature. 
The observations are ON--ON observations, alternating the source
and the sky in each feed horn. A typical integration
time for one flux density data point is 1200--1600 seconds,
and the detection limit under optimal weather conditions 
is approximately 0.2 Jy.
As a primary flux calibrator we use DR21 and as secondary calibrators 3C84 and 3C274. 
The error values include the contribution from the measurement rms and 
the uncertainty of the absolute calibration. 
For more details about the Mets\"ahovi
observing system and data reduction see \citet{terasranta98}.

The observations reported in this paper were performed between years
2001.95 and 2005.27. All flux densities and upper limits are listed in Table~\ref{fluxes}. The number of observations, $N_\mathrm{obs}$, for each source varies greatly. The
most observed sources include BL Lac, OJ 287, Mrk 421, AO 0235+164 and
S5 0716+714. They all have more than 100 data points, S5 0716+714
topping the list with 355 measurements. Such a high number of measurements is exceptional because an overwhelming majority of the sample, 96\%, have
$N_\mathrm{obs}<10$. The source S5 0716+714 was the target of an intensive multifrequency campaign
in 2003 November \citep{ostorero06}. Before the
start of the actual campaign it exhibited extremely active behaviour
in the radio domain and was monitored very frequently especially
in late 2003 and 2004. 

The observation interval for each source varies also; for sources with
$N_\mathrm{obs}>5$ the average interval is 0.34 years, or approximately 124
days. For objects observed 5 times or less, the average
interval is of the order of 13 months.

Some of the best-known, radio-bright sources in the sample have been observed at Mets\"ahovi also prior to 2001 
as part of the extragalactic sources monitoring project \citep{terasranta92, terasranta98, 
terasranta04, terasranta05}. Many of them have frequently-sampled flux curves dating back 
to the beginning of 1980's. If these data are included, 21 sources in all have more than 100 
measurements. We will discuss the flux curves and variability timescales of this limited BL Lac sample 
using the full data set in a forthcoming paper (Nieppola et al. 2007, in preparation).

\begin{figure}[h]
\plotone{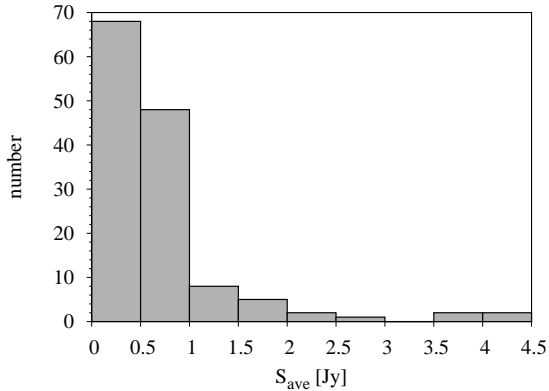}
\caption{The distribution of the average fluxes of the sample sources.\label{s_ave}}
\end{figure}

\begin{figure}[h]
\epsscale{1}
\plotone{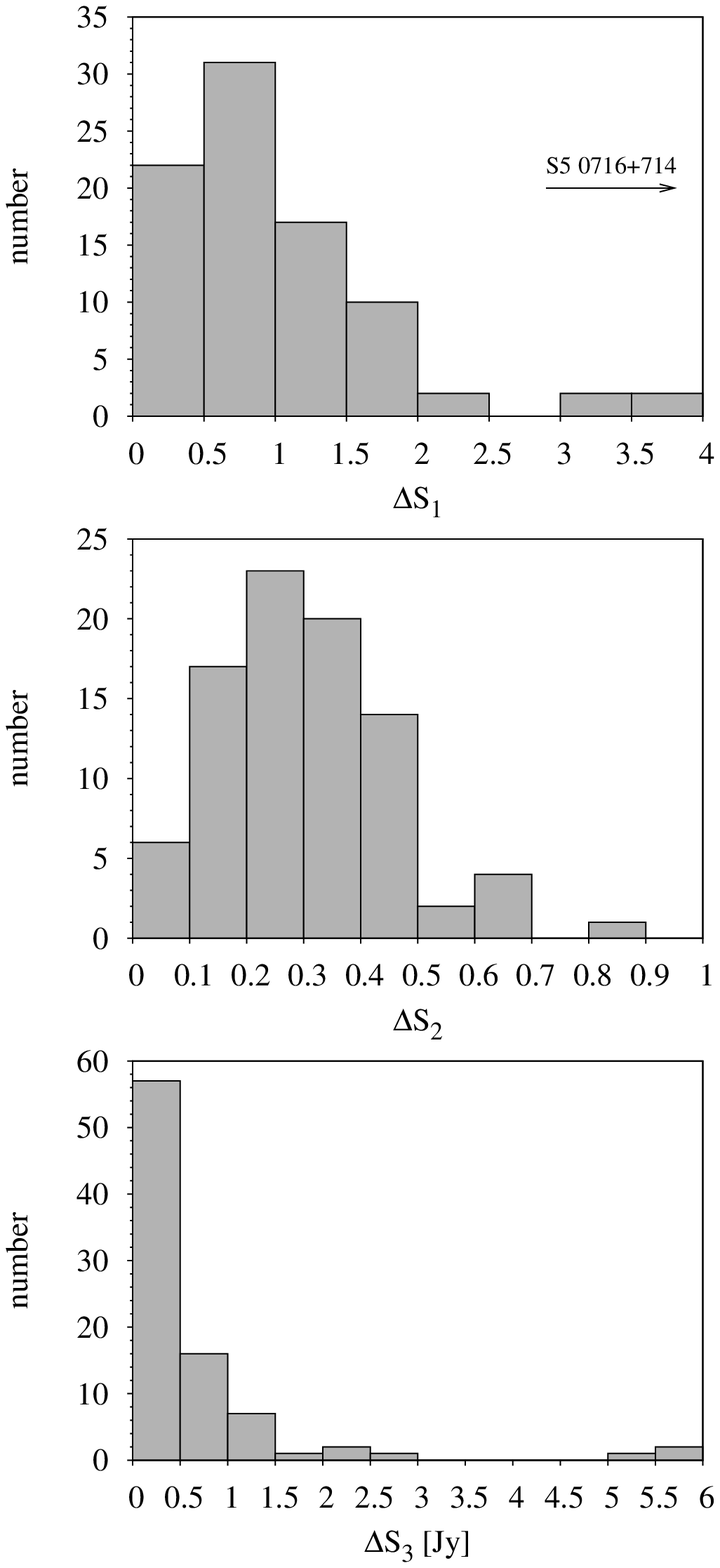}
\caption{The distributions of the variability indices $\Delta S_1$, $\Delta S_2$ and $\Delta S_3$. The object S5 0716+714 ($\Delta S_1$\,=17.47) has been omitted from the upper panel for the sake of clarity.\label{var_hist}}
\end{figure}

\section{DETECTION RATES}
\label{dets}

Table~\ref{detections} lists the detection rates for the whole sample
and the BL Lac classes separately. The detection limit is $S/N>$\,4, typically 
ranging from 0.2 to 0.6 Jy depending on weather conditions. 
Approximately one third of the sample has been detected at $S/N>$\,4 at 37 GHz,
although all of them have been measured at least once. Most of the
detected sources are LBLs, and only one tenth are HBLs. This is due to the larger average radio luminosities of LBLs compared to
HBLs. Luminosity correlations of BL Lac classes are discussed in
detail in \citetalias{nieppola06}. The detected HBLs include objects which were
at first deemed unlikely to be detected. Typically, there are a few
non-detections and one detection. This suggests that there is variability in these sources as well, but normally we cannot detect it due
to the faintness of the source. Fig.~\ref{s_ave} depicts the distribution of the average fluxes of the sample. It emphasizes the difficulties of detecting the sources; roughly half of the detected sources have average fluxes under 0.5 Jy.

\begin{deluxetable}{lcccccc}
\tabletypesize{\scriptsize}
\tablewidth{0pt}
\tablecaption{Detection rates for the Mets\"ahovi sample.\label{detections}}
\tablehead{
\colhead{Class} & \colhead{number of} & \colhead{\% of class} & \colhead{\% of all} & \colhead{number of} & \colhead{\% of class} & \colhead{\% of all}\\
\colhead{} & \colhead{detected sources} & \colhead{detected} & \colhead{detected sources} & \colhead{undetected sources} & \colhead{undetected} & \colhead{undetected sources}
}

\startdata
all sources & 136 & 34.2 & 100 & 262 & 65.8 & 100\\
LBL & 75 & 76.5 & 55.1 & 23 & 23.5 & 8.8\\ 
IBL & 35 & 36.5 & 25.7 & 61 & 63.5 & 23.3\\
HBL & 16 & 14.6 & 11.8 & 94 & 85.5 & 35.9\\
\enddata
\tablecomments{The detection limit is $S/N>$\,4. All detected sources have at least one $S/N>$\,4 measurement. For all sources, $N_\mathrm{obs}\geq1$.}
\end{deluxetable}

\section{VARIABILITY AMPLITUDES}
\label{var}

To examine the variability amplitude of the sample we calculated 3 variability indices:

\begin{equation}\Delta S_1=(S_{max}-S_{min})/S_{min}\end{equation}
\begin{equation}\Delta S_2=(S_{max}-S_{min})/(S_{max}+S_{min})\end{equation}
\begin{equation}\Delta S_3=S_{max}-S_{min}.\end{equation}

The first two describe the fractional variability, whereas the third is simply
the absolute variability of the source. $\Delta S_1$ gives explicitly the proportion of the flux change to the minimum flux. Thus $\Delta S_1$=1 means that the source has increased its flux level to twice its minimum flux density during our observing period. 

Only $S/N>$\,4 detections were used in calculating the indices. The bare minimum of two detections needed to
estimate the 37 GHz variability were available for 87 sources in total, of which 6
were HBLs, 17 IBLs, 58 LBLs, and 6 were unclassified. Fig.~\ref{var_hist} shows the distributions of the indices. The average and median variabilities for the whole sample
and BLO classes separately are listed in Table~\ref{vari}.

\begin{deluxetable}{lcccccccc}
\tablecolumns{9}
\tablewidth{0pt}
\tablecaption{Variability indices for the whole Mets\"ahovi sample and BL Lac
  classes separately. \label{vari}}
\tablehead{\colhead{Class} & \colhead{$\Delta S_1$} & \colhead{$\Delta S_2$} & \colhead{$\Delta S_3$} & \colhead{$\Delta S_1$} & \colhead{$\Delta S_2$} & \colhead{$\Delta S_3$} & \colhead{$N_\mathrm{det}$} & \colhead{$N_\mathrm{det}$}\\
\colhead{} & \colhead{average} & \colhead{average} & \colhead{average} & \colhead{median} & \colhead{median} & \colhead{median} & \colhead{average} & \colhead{median}
}
\startdata
\multicolumn{9}{c}{all data}\\
\tableline
all sources & 1.19 & 0.31 & 0.69 & 0.84 & 0.30 & 0.39 & 18.05 & 5.00 \\
LBL & 1.36 & 0.33 & 0.87 & 0.93 & 0.32 & 0.41 & 21.22 & 6.00\\
IBL & 0.84 & 0.26 & 0.32 & 0.50 & 0.20 & 0.24 & 10.00 & 4.00\\
HBL & 0.78 & 0.26 & 0.32 & 0.83 & 0.30 & 0.34 & 20.67 & 3.00\\
\cutinhead{$N_\mathrm{det}<$\,9}
all sources & 0.85 & 0.27 & 0.55 & 0.72 & 0.26 & 0.38 & 5.41 & 5.00 \\
LBL & 0.93 & 0.29 & 0.68 & 0.83 & 0.29 & 0.39 & 5.74 & 6.00 \\
IBL & 0.70 & 0.23 & 0.28 & 0.50 & 0.20 & 0.20 & 4.88 & 4.00 \\
HBL & 0.52 & 0.20 & 0.24 & 0.50 & 0.20 & 0.27 & 4.67 & 3.00 \\
\enddata
\tablecomments{Indices are as described in \S\ref{var} and $N_\mathrm{det}$ is the
  number of $S/N>$\,4 detections for one source. In the upper half all data is
  included, while in the lower half data is limited to $N_\mathrm{det}<$\,9. See text for details.}
\end{deluxetable}

In the distributions of $\Delta S_1$ and $\Delta S_2$ (Fig.~\ref{var_hist}, upper and middle panels) most of the objects can be found in the lower end of the range, followed by a high-variability tail. A clear exception from the general distribution is S5 0716+714, whose $\Delta S_1$ is anomalously high, 17.47, and $\Delta S_2$=0.9. For this object an unprecedented radio outburst was recorded during
our observing period (see also \S6). The peaks of the distributions are $\Delta S_1$\,=0.5-1.0 and $\Delta S_2$\,=0.2-0.3.   Looking at the lower panel of Fig.~\ref{var_hist}, it is clear that although the fractional variability of the sample sources is relatively high, their absolute variability is not. Even out of the detected sources, a large majority is very faint and the flux levels increase less than 0.5 Jy. Noteworthy are the three sources in the far end of the scale, exhibiting flux changes of more than 5 Jy. These objects are S5 0716+714, OJ 287 and BL Lac.

\begin{figure}
\epsscale{1}
\plotone{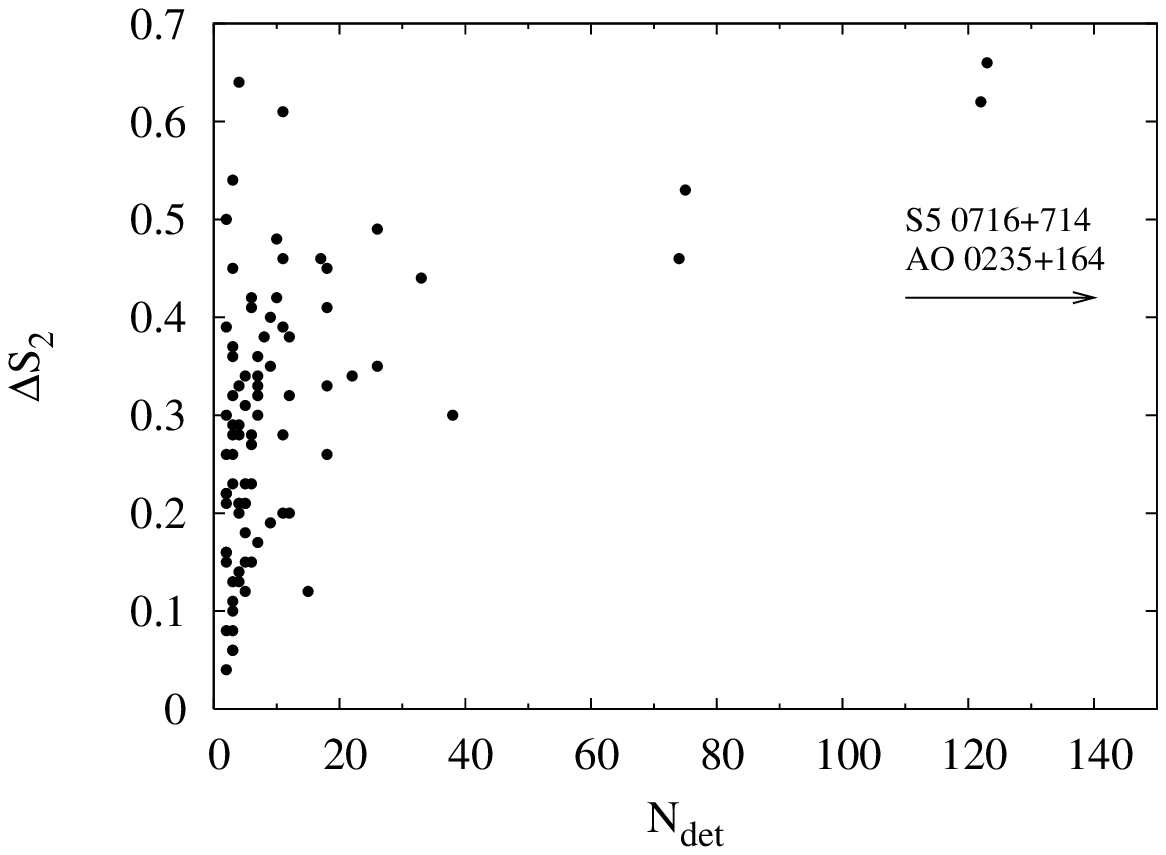}
\caption{Variability $\Delta S_2$ plotted against the number of detections,
 $N_\mathrm{det}$. Data points for S5 0716+714 ($N_\mathrm{det}$=353, $\Delta S_2$=0.9) and AO 0235+164
  ($N_\mathrm{det}$=205, $\Delta S_2$=0.44) have been excluded from the figure for the sake
  of clarity.\label{var_N}}
\end{figure}

As we see from the average results in Table~\ref{vari}, LBLs seem to exhibit slightly stronger variability than
HBLs. The variability of IBLs is approximately the same as that of HBLs. However, there is a significant correlation
between the number of detections, $N_\mathrm{det}$ and variability (Fig.~\ref{var_N}), according to Spearman Rank Correlation test. Radio-bright LBLs are observed more often and
therefore they get a larger $N_\mathrm{det}$. The more measurements there are of a source, the more
probable it is that it is observed both during a quiet and a flaring
state, thus increasing its detected variability. To eliminate this selection
effect, we calculated revised variability parameters for a sample in which
$N_\mathrm{det}<$\,9 for all sources. For those 30 sources that had more than 9 detections, 8 representative points were selected, with as regular time intervals as possible. We chose $N_\mathrm{det}$=8 as the limit of number of data points because below that point the correlation between $N_\mathrm{det}$ and $\Delta S_2$ in Fig.~\ref{var_N} disappears. The results for the $N_\mathrm{det}$-limited sample are also listed in Table~\ref{vari}.

Limiting the value of $N_\mathrm{det}$ affected the two extremes, LBLs and HBLs,
the most, while the variability parameters of IBLs remained almost the
same (Table~\ref{vari}). This is because there are several LBLs with extremely
high $N_\mathrm{det}$ and therefore very high variability, the most extreme object being
S5 0716+714 for which $N_\mathrm{det}$=353. Two examples of
the six HBLs in the sample are the radio-luminous Mrk 421 and Mrk 501
($N_\mathrm{det}$=74 and $N_\mathrm{det}$=38, respectively). Because of the small number of
HBLs, limiting the $N_\mathrm{det}$-values of these two objects had a big impact on
the average variability index of the sample. 

\begin{figure}
\plotone{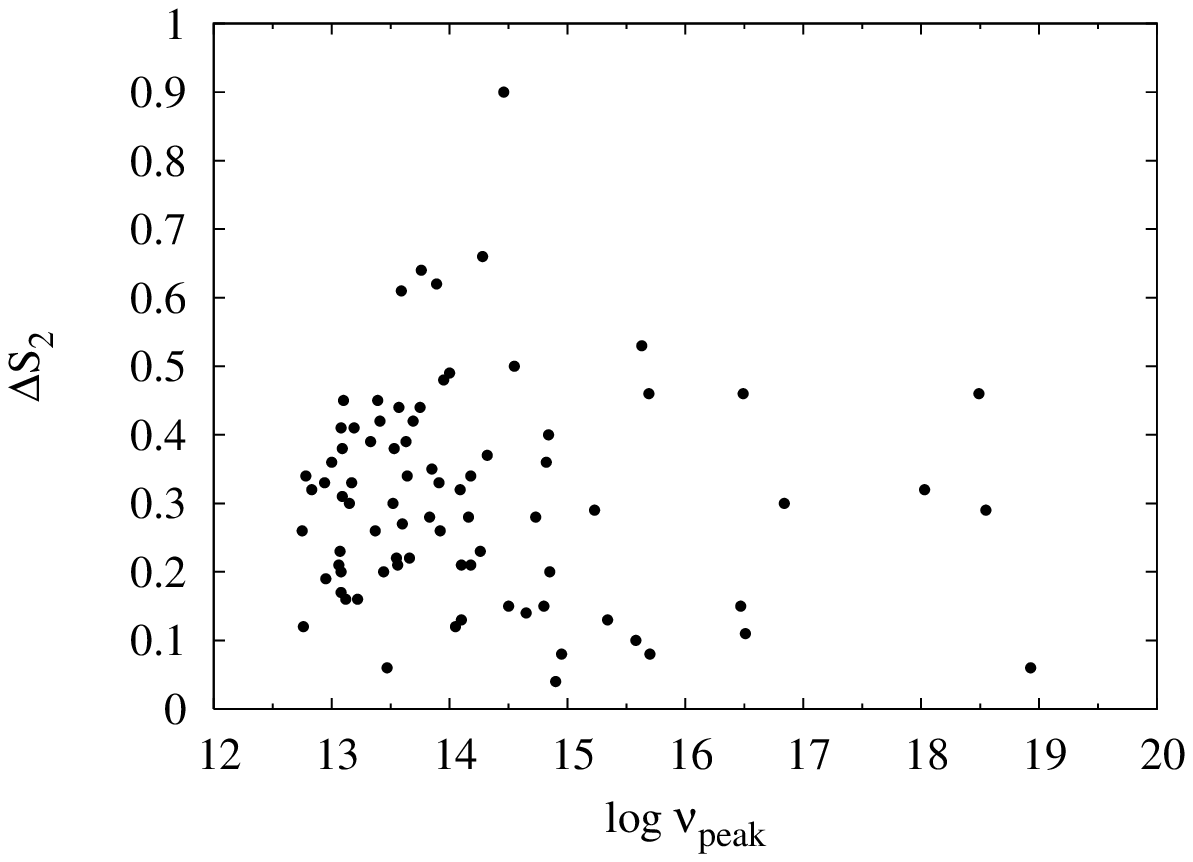}
\caption{Variability $\Delta S_2$ plotted against the synchrotron peak frequency $\nu_{peak}$.\label{corr_peak}}
\end{figure}

 We looked for a possible correlation between $\Delta S_2$ (Fig.~\ref{corr_peak}) and the synchrotron peak frequency $\nu_{peak}$
(as calculated in \citetalias{nieppola06}) using the Spearman Rank Correlation Test. There was
no significant correlation ($P=$0.25). The lack of correlation between $\Delta S_2$ and the synchrotron peak frequency holds also
for the $N_\mathrm{det}$ -limited sample. This further emphasizes the continuity and uniformity of the BLO population found in Paper I. Usually RBLs are said to be more variable than XBLs, but our results so far give no sign of such a distinction.

\section{SPECTRAL INDICES}

\begin{figure*}
\epsscale{2}
\plotone{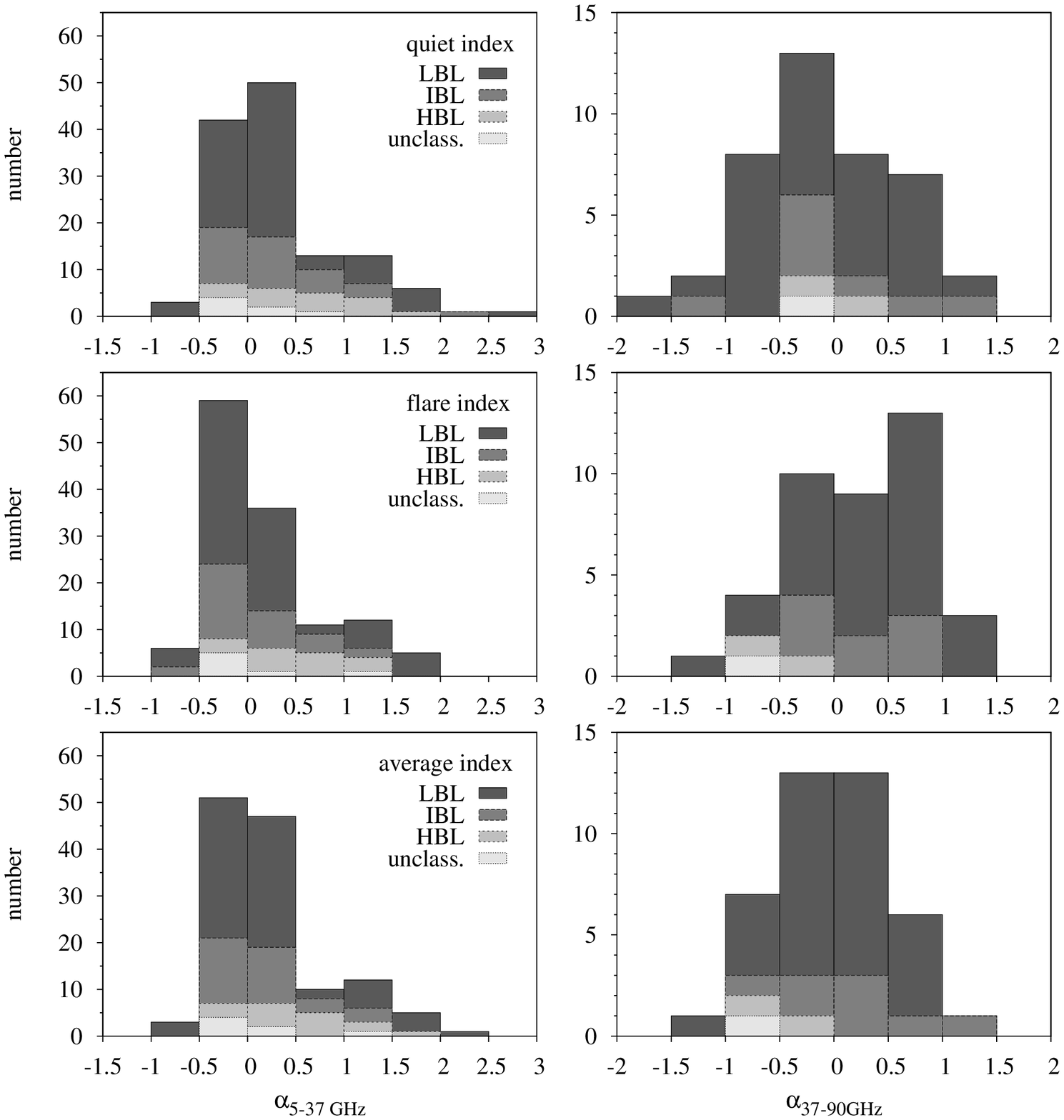}
\caption{Distributions of the quiet, flare and average spectral indices for the sample in the frequency intervals 5 - 37 GHz and 37 - 90 GHz.\label{spec_hist}}
\end{figure*}

We calculated spectral indices for frequency intervals 5\,-\,37 GHz
and 37\,-\,90 GHz to investigate the spectral behaviour of BL Lacs
below and above 37 GHz. 
The spectral index is defined as
$S_\nu\propto\nu^\alpha$. Only $S/N>$\,4
detections were used. The results by class are
shown in Table~\ref{specind}. The 37 - 90 GHz results for HBLs are omitted beacuse there were only 2 HBLs with both 37 and 90 GHz data. The distributions of the indices can be found in Fig.~\ref{spec_hist}. The 5 GHz data are taken from the Astrophysical
Catalogues Support System \citep[CATS\footnote{http://cats.sao.ru}, ][]{verkhodanov97} maintained by the Special Astrophysical
Observatory, Russia. The 90 GHz data 
are taken both from CATS and from our observations with the Swedish-ESO Submillimetre Telescope
(SEST) between 1987 and 2003.

In Table~\ref{specind} and Fig.~\ref{spec_hist} there are 3 different
indices, $\alpha_{quiet}$, $\alpha_{flare}$ and
$\alpha_{ave}$. They are calculated from the minimum
flux densities, maximum flux densities, and average flux densities, respectively. 
In both frequency intervals the range of indices was quite large. The maxima of average indices
were 2.23 and 1.09, and minima --0.56 and --1.47 in 5-37 GHz and 37-90 GHz,
respectively. As we see from Fig.~\ref{spec_hist}, the majority of the sources have flat ($\alpha\,>$\,--\,0.5) spectra in both intervals. The 5 - 37 GHz distribution has a tail towards the higher indices. There are 18 objects with $\alpha_{ave}\,>$\,1. Typically, these sources are X-ray selected BLOs with little radio data. In 37 GHz in particular, 13 sources have only one $S/N>$\,4 datapoint. It is plausible that the source has been in a flaring state at the time of the observation, which would result in an inverted spectrum. The 37 - 90 GHz distribution is less skewed but broader. S5 0716+714 has a conspicuously steep spectrum; its $\alpha_{ave}$\,=-1.47. The steepness is exaggerated by the non-simultaneity of the data; the 37 GHz average flux is strongly influenced by the high flux levels of the 2003 flare, whereas the 90 GHz data is from a quiescent state at an earlier epoch. For the whole sample, most of the data taken at 37 GHz were
from different observing epochs than the 5 GHz or 90 GHz data.
Thus we possess no information of the true shape of the continuum
spectra during the various activity states. For a small subset
of our sample we have managed to obtain truly simultaneous
multifrequency spectra between 1 and 37 GHz. These results will be 
discussed in Tornikoski et al. (in preparation). The non-simultaneous spectral indices calculated in
this paper only give us a crude estimate of the spectral behaviour.

\begin{deluxetable}{lcccccc}
\tablecolumns{7}
\tablewidth{0pt}
\tablecaption{Spectral indices by class for frequency intervals 5 - 37 GHz and 37 - 90 GHz. \label{specind}}
\tablehead{\colhead{Class} & \colhead{$\alpha_{quiet}$} & \colhead{$\alpha_{flare}$} & \colhead{$\alpha_{ave}$} & \colhead{$\alpha_{quiet}$} & \colhead{$\alpha_{flare}$} & \colhead{$\alpha_{ave}$}\\
\colhead{} & \colhead{average} & \colhead{average} & \colhead{average} & \colhead{median} & \colhead{median} & \colhead{median}
}
\startdata
\multicolumn{7}{c}{5-37 GHz}\\
\tableline
all sources & 0.33 & 0.17 & 0.25 & 0.16 & 0.00 & 0.05\\
LBL & 0.30 & 0.15 & 0.23 & 0.15 & -0.02 & 0.01\\
IBL & 0.28 & 0.07 & 0.16 & 0.13 & -0.02 & 0.02\\
HBL & 0.67 & 0.51 & 0.60 & 0.68 & 0.55 & 0.51\\
\cutinhead{37-90 GHz}
all sources & -0.10 & 0.15 & 0.00 & -0.12 & 0.22 & -0.02\\
LBL & -0.13 & 0.21 & 0.01 & -0.10 & 0.36 & -0.01\\
IBL & -0.03 & 0.23 & 0.09 & -0.18 & 0.22 & 0.13\\
\enddata
\tablecomments{ 37-90 GHz indices of HBLs are omitted due to small sample size. The indices have been calculated with non-simultaneous archival data.}
\end{deluxetable}

There are no major differences in the spectral indices between the BLO classes among the detected sources. For HBLs the 5-37 GHz indices are 
slightly more inverted than for LBLs and IBLs. Spearman Rank Correlation
test found no correlation between the spectral indices and the synchrotron peak
frequency $\nu_{peak}$, except for a marginal positive correlation for $\alpha_{flare}$ and $\alpha_{ave}$ in 5\,-\,37 GHz. 

Generally, there is a correlation between spectral index and variability,
steep spectrum sources being on average less variable than flat spectrum
sources \citep{edelson87, valtaoja92_moniIV}. However, among the flat spectrum sources, and especially among BL Lacs,
no significant correlation has been detected. Our
study makes no exception: the 37 GHz variability depends on neither of the spectral
indices according to the Spearman Rank Correlation test.

\section{TIMESCALES AND FLUX CURVES}

While all detected BL Lacs seem to exhibit some degree of
variability, the timescales are diverse. We calculated the slopes of
the flux density changes for each source. S5 0716+714 has clearly the fastest
flux density variations. It has exhibited documented intraday
variability (IDV) behaviour, and has also been a subject of a very
intensive monitoring campaign in 2003.
Its flux density increased by 1.06 Jy in 40
minutes and 1.12 Jy in less than 2 hours, and decreased by 0.87 Jy in less than 2 hours during the 2003 flare. The multifrequency
flare and the IDV during the campaign are discussed in detail
in \citet{ostorero06}.

AO 0235+164 has the second fastest variations: its flux density increased by approximately 0.4 Jy in less than an hour in 2004 August, after dropping from 2.38 to 1.76 Jy in 40 minutes 16 hours
earlier. Also OJ 287 is worth mentioning with its flux density increase of 0.8
Jy in approximately 2 hours. 

There were 5 sources, which had a well-sampled flare during our observing period: AO 0235+164, S5 0716+714, OJ 287, 1308+326, and BL Lac, all of them LBLs. No two flares were alike; their type ranged from powerful and short (S5 0716+714, 2003 July - 2003 December, peak flux density 6.28 Jy) to broad and weak (1308+326, 2002 January - 2004 October, peak flux density 3.28 Jy). OJ 287 had two different outbursts. The later one in 2003 August - 2005 April reached the highest flux density value recorded in our observations, 7.8 Jy. The flare of BL Lac was still ongoing in 2005 April. The flux density had risen steadily since 2004 February, and had reached 6.6 Jy in 2005 March. The mean duration of a flare is 17.6 months. 

Most of the sources in our sample have very few data points and thus no
flux curve to speak of. As mentioned in \S\ref{dets}, for many of these faint sources --
mainly HBLs and XBLs -- our data consist of detections (S/N$>4$) and
non-detections (S/N$<4$). This is partly due to the varying observing
conditions, with the detection limit ranging from 0.2 to
over 0.5 Jy, but it can also, at least partially, be caused by variability. 
There are examples of sources, expected to be faint and usually below the detection limit, suddenly
measuring over 0.5 Jy, such as MS 0737.9+7441 and
GB 1011+496. An especially good
example of such behaviour is RXS J1110.6+7133. It has one detection at 0.74 Jy and 6
non-detections.

\section{DISCUSSION}

In this paper we publish the data of the first 3.5 years of our ongoing BL Lacertae object observing project. We have also calculated some variability parameters and spectral indices for the sample.

The sample consists of 398 BL Lacs. During the 3.5 years of
monitoring we made at least one significant ($S/N >$ 4) detection of slightly
more than one third of the sources at 37 GHz, with the detection
limit being of the order of 0.2--0.5\,Jy. The largest subpopulation
that was detected was LBLs (77\%). 37\% of IBLs and
15\% of HBLs were detected. Even though the percentage of the
detected BLOs is smaller in the high-energy peaked population
than in the low-energy peaked one, it is interesting to see
that also among them there is a large number of objects that
were detected at 37 GHz.
This has a significant implication for the extragalactic foreground point source modelling of the Planck satellite. Because the Planck 
detection limit will be comparable to that of the
Mets\"ahovi telescope, approximately one third the BLO population could 
also be detected by Planck. This should include at least
some of the high-energy peaked BLOs that are often excluded from both high-frequency radio observing campaigns as well as many of
the models for CMB contribution by blazars.
\citet{giommi04} suggested that blazar flux variability at
millimetre wavelengths may be a significant contributor to the CMB
map contamination.
Our findings of variability within the BLO sample, including clear signs of
variability for sources in the HBL population, confirm this.

In order to have a reasonable estimate of how probable it is
to observe the faint BLOs above the detection limit,
we calculated the ratio of detections versus non-detections.
For the HBLs for which there is at least one detection in our data,
but excluding the two densely monitored and usually detected
objects Mrk 421 and Mrk 501, the percentage of detections among
all good quality data points was 19\%.
If we add the X-ray selected BLOs from the IBL and LBL subclasses, the
percentage is 22\%. Thus the probability to detect an HBL or XBL at a random 
observing epoch is of the order of 20\%.
This either means that their flux density remains on the verge of our
weather-dependent detection limit, or that they are variable, flaring
sources. 

Our earlier studies on Southern flat-spectrum radio sources \citep{tornikoski00_southern} show that many radio-bright AGNs
spend much less time in an active state than in a quiescent or
intermediate state. Our long term monitoring project of radio-bright
AGNs \citep[][ Hovatta et al. 2006, in preparation]{terasranta05} shows that 3.5 years of data is a rather short time to make any
conclusive statements about such variability
behaviour as typical peak fluxes of flares or peak--to--peak timescales, because many of the dramatic changes in the flux density
occur on long timescales. Thus it is important to consider the
possible effects of long term variability when interpreting
results of a few years' monitoring campaign. While we can
say that the BLOs detected during this campaign
can be bright in the mm-domain, we cannot rule out the 
possibility that the ones that now remained undetected 
could be much brighter if observed in a flaring state.
This is an important issue when considering the contamination of CMB maps caused by extragalactic foreground sources, the data used
for the modelling are one to few-epoch historical data.
We will discuss the long term variability of a subsample of these BLOs 
in more detail in Nieppola et al. (2006, in preparation).

The variability indices $\Delta S_2$ of our sample range from 0.04 to 0.9. The brightest sources are observed more often whereas with the faint ones we make do 
with just a couple of measurements. This creates a selection effect and leads to biased variability behaviour, 
as explained in \S\ref{var}. For instance, in \citet{ciaramella04} the 37 GHz variability index for LBLs 
was $\Delta S_2$=0.74\,$\pm$\,0.27. Our average value was $\Delta S_2$=0.34. Our value is lower because it includes 
many rarely observed faint LBLs. Therefore it is very problematic to assign any variability 
value to a source, much less a source class, as it is always case-specific. Considering the differences in the results of the complete sample and 
the $N_\mathrm{det}$ -limited one, we find that the variability index $\Delta S_2$ changes little compared to $\Delta S_1$ and 
$\Delta S_3$. This is because, as the difference between $S_{max}$ and $S_{min}$ increases, $\Delta S_2$ approaches 
unity. Meanwhile, $\Delta S_1$ grows without a limit and $\Delta S_3$ approaches $S_{max}$. Therefore $\Delta S_2$ 
seems to be least sensitive to the effect of a varying number of measurements, and as such probably the most convenient description of variability.

When we compare the variability characteristics of the BLO classes -- LBLs, IBLs and HBLs -- no definite differences 
can be found. This is unexpected. Radio-selected BL Lacs (RBLs) are generally thought to be on average more variable 
than X-ray selected BL Lacs (XBLs). Our results indicate that this is not necessarily true for LBLs and HBLs. However, we have to keep in mind that we can only study the 
radio-bright HBLs, whose properties one could expect to be similar to those of LBLs. Over 80 \% of HBLs still 
remain undetected at 37 GHz. 

The 5 - 37 GHz and 37 - 90 GHz spectral indices of the sample are diverse. In the lower frequency band the spectra are mostly flat, as expected, but in the higher band the distribution broadens to include both rising and falling spectra. This indicates that although BLOs are flat spectrum sources at low frequencies, in the high radio frequencies their spectral behaviour is not necessarily uniform. 

The high-index tail of the 5-37 GHz spectral index distribution includes mostly XBLs with only one detection at 37 GHz. This could be thought of as further evidence of the flaring nature of the faint BLO population. If we assume that in quiescence, these XBLs have similar spectra to those of the vast majority of BLOs, their quiet flux level would be roughly of the order of 50 mJy or less. Thus, to be detected at Mets\"ahovi, their flux densities must increase four-fold at the very least. We believe that active monitoring of the faint BLO population would result in more such detections. Also, using a more sensitive observing system to study a limited sample would be fruitful.

\section{CONCLUSIONS}

We have observed a sample of 398 BL Lacertae objects in Mets\"ahovi Radio Observatory from 2001 December to 2005 April. 
We obtained a $S/N>\,$4 detection for 34\% of the sources. The detected fraction of LBLs was 77\%, of IBLs 37\% and 
of HBLs 15\%. These figures are concordant with the typical radio-faintness of HBLs, as observed in \citet{nieppola06}. 
However, we stress that HBLs cannot be considered radio-silent; this is verified by the serendipitous detections 
we obtained for some of them. 

\acknowledgments
We gratefully acknowledge the funding from the Academy of Finland for
our Mets\"ahovi and SEST observing projects (project numbers 205969, 46341, and
51436). We acknowledge EC funding under contract HPRN-CT-2002-00321 (ENIGMA). The authors made use of the database
CATS \citep{verkhodanov97} of the Special Astrophysical Observatory.

Planck satellite is a mission of the European Space Agency ESA. Our work
has been carried out within its Low Frequency Instrument Consortium
Working Group 6 (Extragalactic foreground sources).

{\it Facilities:} \facility{Mets\"ahovi Radio Observatory ()}, \facility{SEST (), \facility{CATS}}

\clearpage

\clearpage

\end{document}